# Signature of Topological Semimetal in Harmonic-honeycomb ReO$_3$

*Yifeng Han[1,#], Cui-Qun Chen[2,#], Hualei Sun[3], Shuang Zhao[1], Long Jiang[4], Yuxuan Liu[1], Zhongxiong Sun[1], Meng Wang[5], Hongliang Dong[6], Ziyou Zhang[6], Zhiqiang Chen[6], Bin Chen[6], Dao-Xin Yao[2,\*], Man-Rong Li[1,7,\*]*

[1]Key Laboratory of Bioinorganic and Synthetic Chemistry of Ministry of Education, School of Chemistry, Sun Yat-sen University, Guangzhou 510006, P. R. China.

[2]Guangdong Provincial Key Laboratory of Magnetoelectric Physics and Devices, State Key Laboratory of Optoelectronic Materials and Technologies, School of Physics, Sun Yat-sen University, Guangzhou, 510275 P. R. China

[3] School of Science, Sun Yat-sen University, Shenzhen, Guangdong 518107 P. R. China.

[4]Instrumental Analysis & Research Center, Sun Yat-sen University, Guangzhou, 510275 P. R. China

[5]Center for Neutron Science and Technology, Guangdong Provincial Key Laboratory of Magnetoelectric Physics and Devices, School of Physics, Sun Yat-Sen University, Guangzhou, Guangdong 510275 P. R. China

[6]Center for High Pressure Science and Technology Advanced Research, Shanghai 201203 P. R. China

[7]School of Chemistry and Chemical Engineering, Hainan University, Haikou, 570228 P. R. China

Email: limanrong@mail.sysu.edu.cn (M.-R. Li), yaodaox@mail.sysu.edu.cn (D.-X. Yao)

# These two authors contributed equally to this work.

# ABSTRACT

Transition-metal honeycomb compounds are capturing scientific attention due to their distinctive electronic configurations, underscored by the triangular-lattice spin-orbit coupling and competition between multiple interactions, paving the way for potential manifestations of phenomena such as Dirac semimetal, superconductivity, and quantum spin liquid states. These compounds can undergo discernible pressure-induced alterations in their crystallographic and electronic paradigms, as exemplified by our high-pressure (HP) synthesis and exploration of the honeycomb polymorph of $ReO_3$ ($P6_322$). This HP-$P6_322$ polymorph bears a phase transition from $P6_322$ to $P6_3/mmc$ upon cooling around $T_p = 250$ K, as evidenced by the evolution of temperature-dependent magnetization (*M-T* curves*)*, cell dimension and conductivity initiated by an inherent bifurcation of the oxygen position in the *ab* plane. Insightful analysis of its band structure positions suggests this HP-$P6_322$ polymorph being a plausible candidate for Dirac semimetal properties. This phase transition evokes anomalies in the temperature-dependent variation of paramagnetism (non-linearity) and a crossover from semiconductor to temperature-independent metal, showing a temperature independent conductivity behavior below ~200 K. Under increasing external pressure, both the $T_p$ and resistance of this HP-polymorph are slightly magnetic-field dependent and undergo a "V"-style evolution (decreasing and then increasing) before becoming pressure independent up to 20.2 GPa. Theoretical calculations pinpoint this anionic disorder as a probable catalyst for the decrement in the conductive efficiency and muted temperature-dependent conductivity response.

# 1. Introduction

Honeycomb layered materials has been extensively studied in recent years, due to their exceptional mechanical, electrical, and optical properties for promising applications in energy storage and quantum devices.[1-3] Honeycomb, hyper-honeycomb, and harmonic-honeycomb materials are all categorized by their hexagonal lattice structure,[4-6] where the transition-metal enriched materials are subjected to spin frustration, creating competition between interactions and leading to multiple magnetic configurations to satisfy the principle of minimum energy within a singular state.[7, 8] The resulting exchange frustration, stemming from interlinked spins of each elementary triangle, contributes to not only the emergence of distinct physical features such as quantum spin-liquid (QSL), Dirac semimetal, and high temperature superconductivity, but also the observed semi-metallic band structure.[9-14]  Particularly, in the paramagnetic state with half-filled $e_g$ orbitals, Dirac cones can arise due to electronic transitions in the inherent honeycomb superlattice. These intricacies make such sophisticated layered structures an intriguing platform for probing correlated Dirac electrons and exploring topologically nontrivial magnetic phenomena. Honeycomb layered structure, consisting of two different layers alternately stacked along the $c$-axis (**Fig. 1a**), exhibits a high degree of stacker patterns and topological properties, making them a fundamental building block for hexagonal compounds. $Na_2IrO_3$

is believed to be a promising candidate for realizing Kitaev physics, since its temperature-dependent susceptibility ($\chi(T)$) data indicate a transition into long-range antiferromagnetically ordered state below the Néel temperature ($T_N$ = 15 K), portraying a type of topological quantum magnetism with valuable applications in quantum computing.[15] Li$_2$RuO$_3$, on the other hand, exhibits an interesting honeycomb lattice distortion and possesses robust spin-orbit coupling (SOC), leading to a metal-insulator transition (MIT) at $T_C$ = 540 K, making it a promising candidate for studying exotic magnetic phenomena.[16] Li$_2$MnO$_3$ displays antiferromagnetic (AFM) ordering below $T_N$ of 36 K, characterized by an ordered magnetic moment of 2.3 $\mu_B$ oriented perpendicular to the *ab* plane. Notably, 35% of the total magnetic entropy is released in the seemingly simple paramagnetic phase in Li$_2$MnO$_3$, which suggests the presence of typical spin dynamics at elevated temperatures. These unique magnetic properties render Li$_2$MnO$_3$ a promising candidate for spintronics and magnetics research.[17] CaHS$_3$ with two-dimensional HS and CaH$_2$ sheets is metastable and exhibits metallic behavior above 128 GPa. The presence of van Hove singularities in CaHS$_3$ increases its density of states (DOSs) at the Fermi level and consequently raises its superconducting critical temperature. This temperature is estimated to be as high as ~100 K at 128 GPa.[18]

Hyper-honeycomb compounds remain a three-dimensional (3D) extension of honeycomb layered lattice, where each hexagonal lattice has four neighbor ones as shown in **Fig. 1b**, leading to a more complex structure and topological properties. The hyper-honeycomb $\beta$-Li$_2$IrO$_3$, which exhibits a zigzag AFM ground state, serves as an excellent example of the potential realization of Kitaev physics.[19] In addition, its $\chi(T)$ curves confirms the existence of bond-dependent ferromagnetic (FM) interactions between the $J_{\text{eff}}$ = 1/2 moments, as evidenced by the positive Curie-Weiss temperature $\theta_{\text{CW}} \sim$ 40 K. Magnetic field can induce a significant moment greater than 0.35 μB per Ir ion at 3 T or higher in an ordered state, resulting in a highly polarized $J_{\text{eff}}$ = 1/2 state in $\beta$-Li$_2$IrO$_3$, which is critical in developing innovative electronic instruments with extremely low power consumption.[19] Na$_4$Ir$_3$O$_8$ is another hyper-honeycomb example exhibiting unusual magnetization behavior. It contains an antiferromagnetically coupled $S$ = 1/2 spin system formed on a geometrically frustrated hyperkagome lattice. Magnetization and magnetic specific heat data show the absence of long-range magnetic ordering at least down to 2 K in Na$_4$Ir$_3$O$_8$. These findings highlight the enormous potential of Na$_4$Ir$_3$O$_8$ for applications in quantum computing, spintronics, and other fields of condensed matter physics.[20]

Harmonic-honeycomb materials typically possess symmetrical cell properties as depicted in **Fig. 1c**. The hexagonal arrangement of atoms in this kind of hexagonal lattice allows each atom to vibrate in three directions conducted in a tuned-sum fashion and governed by the lattice symmetry. These materials usually exhibit low thermal conductivity and high coefficient of thermal expansion, being thermoelectric and thermally stable. Particularly, the unique crystal structure and harmonic vibrational properties in harmonic-honeycomb lattice render them characteristics similar to those of phonon crystals in condensed matter physics for phonon transport and phonon optics applications.[5] The representative $\gamma$-Li$_2$IrO$_3$ exhibits strong SOC with competing magnetic interactions into an incommensurate spiral spin order at ambient pressure below 38 K, making it a potential candidate for realizing Kitaev physics. Recent studies have suggested that applying magnetic fields to $\gamma$-Li$_2$IrO$_3$ can induce a quantum phase transition from a topological to a non-topological phase.[21]

The structures and properties of those honeycomb layered materials can be further tuned by high-pressure (HP) synthesis and/or medication.[2, 22-25] $\alpha$-RuCl$_3$ exhibits a pressure-induced phase transition from a monoclinic ($C2/m$) to a triclinic ($P$-1) structure.[26] The HP phase exhibits notably robust AFM isotropic bonds, attributable to the heightened direct overlap of the Ru 4d $t_{2g}$ orbitals. The Ru-Ru dimerization completely suppresses the magnetization, resulting in a pressure-induced nonmagnetic state of $\alpha$-RuCl$_3$. Interestingly, $\alpha$-RuCl$_3$ ($C2/m$) demonstrates the occurrence of both a QSL state under magnetic field and a spin solid state under hydrostatic pressure, forming a spin singlet valence bond crystal.[26] The QSL candidate NaYbSe$_2$ demonstrates a pressure-induced MIT at 58.9 GPa and superconductivity emerging at much higher pressure of $P$ = 103.4 GPa, accompanied by a structural phase transition from the $R$-3$mH$ to $P$-3$m$1 symmetry around 11 GPa upon pressing. Moreover, the temperature-dependent resistivity in the metallic state of NaYbSe$_2$ exhibits a crossover from non-Fermi liquid to Fermi liquid behavior at low temperatures. These discoveries provide a promising avenue to investigate the characteristics of MIT and the interplay between spin and charge degrees of freedom in QSL systems with strong SOC.[27]

So far, extensive research has been conducted on honeycomb QSL candidates ($S$ = 1/2)

with strong SOC, albeit with a limited focus on the studies of $Ir^{4+}$ and $Ru^{3+}$ compounds. In contrast, there exists a significant knowledge gap to the exploration of $d^1$-$Re^{6+}$ ($S = 1/2$) honeycomb lattice, which could further enhance our understanding of this unique state. Notably, $Re^{6+}$ offers a contrasting counterpart of $Ir^{4+}/Ru^{3+}$ with regards to the electron hole symmetry. Its occurrence within the $ReO_6$ octahedron and consequent possession of one $5d$ electron in $t_{2g}$ orbital presents an intriguing prospect. In contrast, $Ir^{4+}/Ru^{3+}$ in the $Ir/RuO_6$ octahedron presents one hole.[28, 29] The opposite sign SOC coefficients in $Re^{6+}$ and $Ir^{4+}/Ru^{3+}$ further underscores their differing characteristics. The simplest $Re^{6+}$ oxide, $ReO_3$, adopts an *A* site-deficient perovskite structure (*Pm*-3*m*) with rock-salt ordering $ReO_6$ octahedra. With gradual increase of pressure and temperature in the phase diagram, $ReO_3$ undergoes subsequent structural transitions: *Pm*-3*m*, *Im*-3*m*, *R*-3*c* and $P6_322$ (**Fig. 2a**), accompanied with new physical properties: The *Pm*-3*m* phase, observed under ambient pressure resulting in metallic conductivity.[30, 31] Driven by phonons, this phase exhibits negative thermal expansion of different orders.[32, 33] When an external magnetic field is applied, the conductivity behavior of the system undergoes a transition from metallic to semi-metallic at low temperatures, while its magnetoresistance exhibits a quadratic relationship with the external field.[34] The *Im*-3*m* phase of $ReO_3$ is reached under moderate high pressure, which maintains cubic symmetry but with modified Re-O bond lengths, potentially reducing its metallic behavior and increasing density. At higher pressures, the *R*-3*c* phase emerges, featuring a rhombohedral structure. This phase exhibits distorted Re-O octahedral coordination, and the superconducting temperature is about 17

K at a pressure of 30 GPa. [35] In this paper, we synthesized the hexagonal $P6_322$ phase ReO$_3$ (hereafter denoted as HP-ReO$_3$) by high-pressure and high-temperature (HPHT) synthesis. Our investigations suggest that this HP-ReO$_3$ could serve as a potential Dirac semimetal. *In-situ* pressure- and temperature-dependent structural and physical properties of this $P6_322$ polymorph were also extensively investigated, indicating temperature-dependent oxygen defect evolution in the harmonic-honeycomb lattice, accompanied by pressure-dependent variation of the defect-transition temperature ($T_p$) and resistance.

## 2. Experimental procedures

**2.1 Synthesis and Phase Analysis.** Single crystals of HP-ReO$_3$ (~ 0.27 mm) were prepared by treatment of ReO$_3$ (99.99%, Alfa Aesar) at 1173 K under 8 GPa for 30 min, in which samples were quenched by turning off the power to the heater and slowly depressurized over several hours. The phase purity was examined by laboratory powder X-ray diffraction (PXD) data collected in a powder X-ray diffractometer (MiniFlex 600, Rigaku, Japan) equipped with Cu K$\alpha$ tube ($\lambda$ = 1.5418 Å at 40 kV and 15 mA). The single crystal X-ray diffraction data of HP-ReO$_3$ was collected on an Agilent SuperNova diffractometer using Mo K$a$ ($\lambda$ = 0.71069 Å) micro-focus X-ray sources at 100-300 K.

**2.2 *In Situ* Variable-Temperature PXD Measurements.** *In situ* variable temperature PXD (VT-PXD) data were collected at a Rigaku SmartLab Cu K$\alpha$ tube ($\lambda$ = 1.5418 Å at 40 kV and 40 mA). The polycrystalline powder was loaded into a vacuum sealed glass capillary (diameter ~ 1 mm) and measured at 88, 100, 120, 150 ,170, 180, 190 K and 210-280 K in steps of 10 K. Each scan was collected with $2\theta$ between 10º-60º for approximately 30 min, for which a relaxation time of 5 min was applied at each temperature point before data collection, to extract the lattice parameter evolution upon cooling.

**2.3 Physical Properties Measurements.** Magnetization measurements were carried out with a commercial Quantum Design superconducting quantum interference device (SQUID) magnetometer. The susceptibility was measured in zero field cooled (ZFC) and field cooled (FC) modes under applied magnetic field of 0.1 T between 5 and 300 K. Isothermal magnetization curves were obtained at $T$ = 5-300 K under an applied magnetic field from -5 to 5 T. Ambient pressure electrical resistivity was measured using a physical properties measurement system (PPMS). The *in situ* high-pressure electrical resistance measurements on HP-ReO$_3$ single crystals were carried out using a miniature diamond anvil cell (DAC) made from Be-Cu alloy. Diamond anvils with a 400-μm culet were used, and the corresponding sample chamber with a diameter of 150 μm was made in the insulating gasket. The gasket was achieved by a thin layer of cubic boron nitride and epoxy mixture. NaCl powders were employed as the pressure-transmitting medium. Pressure was calibrated using the ruby fluorescence shift at room temperature. The standard four-probe technique was adopted in these measurements. A micro-Raman system (Renishaw, UK) with 532 nm laser excitation, including 10% laser intensity with 10 s of integration time, was applied to obtain the Raman spectra. The DAC had a pair of 300 μm diamonds with a 300 μm culet size with a T301 steel gasket. Silicone oil was used as a pressure transmitting media.

**2.4 Theoretical Calculations.** Our density functional theory (DFT) calculations were performed utilizing the projector augmented wave (PAW)[36, 37] method within the framework of generalized-gradient approximation (GGA) proposed by Perdew, Burke, and Ernzerhof (PBE), [38] as implemented in the Vienna *ab initio* simulation package (VASP). [39, 40] The energy cutoff for the plane-wave expansion was chosen to be 400 eV. We adopted a $\Gamma$-centered 8 × 8 × 7 Monkhorst Pack *k*-mesh grid to calculate the electronic properties. The experimentally measured lattice constants at different temperature were used in our structural relaxations and only the inner atomic positions were relaxed until the force acting on each atom was smaller than 0.01 eV/Å. The criterion for total energy convergence was set to $10^{-6}$ eV.

Our first-principles calculations for transport properties were carried out separately based on DFT. The VASP with GGA was employed.[38-40] A 3 × 3 × 3 supercell was employed to perform the calculations. In the *ab* plane disordered case, we constructed 64 3 × 3 × 1 supercells at first. Due to the splitting occupancy of O atoms in each unit cell, we made the O atoms occupying *12i* position randomly but kept the ratio of Re and O in each unit cell to be 1:3. Therefore, the O atoms in ab plane should be disordered. Then we staked the 3 × 3 × 1 supercells along the *c* axis. Consequently, we obtained structures which are disordered in the *ab* plane but ordered in the *c* axis. In the other disordered case, 64 3 × 3 × 3 supercells were built at beginning, and the rest steps are the same. The variation of electronic conductivities with temperature of each case are taken from the average values of 64 supercells. The K-mesh was set to 0.02 × 2π /Å. VASPKIT was utilized to analyze the electronic conductivity.[41]

## 3. Results and discussion

### *3.1 Crystal Structure of HP-ReO$_3$*

*In situ low-temperature single crystal diffraction studies*

HPHT syntheses (1173 K and 8 GPa) of HP-ReO$_3$ yield shiny metallic copper-like single crystals (**Fig. 2b** inset) in strong contrast to the red cubic ReO$_3$ precursor. The room-temperature PXD data (**Fig. 2b**) confirmed that the HP produced can be well indexed as the hexagonal HP-ReO$_3$ polymorph (*P*6$_3$22, *a* ~ 4.85 Å and *c* ~ 4.48 Å), in good agreement with reported values. [34,35] The crystallographic data from single crystal diffraction were used to depict the crystal structure of HP-ReO$_3$, given the higher resolution of single crystal diffraction. The final refined crystallographic parameters, agreement factors and selected interatomic distances and bond valence sums (BVS) calculations are listed in **Table S1.** The vertex-sharing ReO$_6$ octahedra are arranged zigzag along the *c*-axis with average bond length <Re-O> = 1.906 (4) Å in HP-ReO$_3$. The ReO$_6$ octahedra form a honeycomb configuration when viewed along the *c*-axis, in strong contrast to the cubic ReO$_3$ phase with <Re-O> = 1.8735 (3) Å.

The low temperature structure of HP-ReO$_3$ was solved by single-crystal X-ray diffraction method as presented in **Table S1**. Below the transition temperature ($T_p$, the phase transition temperature) around 250 K, the crystal structure changes from *P*6$_3$22 to $P6_3/mmc$ (**Fig. 3**), where Re occupies the 2$d$ position and O partially located at the 12$i$ position. The crystallographic data collected at 200 K were selected for more detailed discussion. At 200 K, $ReO_6$ changes from a 6-coordinated octahedron to a pseudo-12-coordinated decahedron, and the edge length of the hexagonal tunnel changes from 1.746 to 1.753 Å viewed alone the $c$-axis, andthe interatomic distance between <Re-Re> increases from 3.5703(4) to 3.5792(4) Å, while the <Re-O> bond length remains the same. Interestingly, the <Re-O> (1.906(4) and 1.910(4) Å) bond length remains almost temperature independent with decreasing temperature. Owing to the splitting occupancy of O atoms, the connection between $ReO_6$ polyhedral changed from vertex-shared to edge-shared arrangement, which should be responsible for the metal-semiconductor transition discussed later. **Fig. S1a** shows the crystal structure of HP-$ReO_3$, where the brown balls illustrate the $Q$ peak, suggesting a weak intensity of electron density in the center of the honeycomb. Structurally, $ReO_3$ can be regarded as a $A$-site vacancy $(\square)_A(Re)_BO_3$ perovskite as previously reported in $Li_2ReO_3$ and $H_xReO_3$.[42, 43] Here we attribute the $Q$ peak (electron density) to the migration of a small amount of $B$-site Re to the $A$-site to form $Re_xRe_{1-x}O_3$ during the HPHT preparation process. The $Q$ peak appeared in the middle of the honeycomb after the temperature quenching (**Fig. S1b**), which may be caused by local inhomogeneity of the honeycomb shape due to O split in the $ab$ plane.

*In Situ VT-PXD characterization*

To systematically track the changes in cell evolution of this HP-$ReO_3$ at low temperatures, the cell parameters extracted from *in situ* VT-PXD data between 280-88 K were plotted in

**Fig. S2.** The values of *a* and *V* decrease linearly with decreasing temperature, in contrast, the *c*-axis shows a small abrupt change at 220 K, which can be attributed to the structural transition from *P*6$_3$22 to *P*6$_3$/*mmc*. In the quoted temperature interval, the volumetric contraction from 92.105(3) to 92.018(4) Å$^3$ with an average *V*-based thermal expansion coefficient ($\alpha = \frac{1}{V_0}\left(\frac{\partial V}{\partial T}\right)$) of 0.5 × 10$^{-5}$ K$^{-1}$ (**Fig. S2**). These values are close to those in the zero thermal expansion materials such as 0.7PbTiO$_3$-0.3BaTiO$_3$ (-0.6 × 10$^{-5}$ K$^{-1}$).[44]

## *3.2 Physical Properties of HP-ReO$_3$*

*Electrical conductivity*

Temperature-dependent four-probe *dc* conductivity measurements (**Fig. 4a**) were initially conducted on HP-ReO$_3$ crystals at ambient pressure, ranging from 2 to 300 K. The recorded data reveal a sharp resistivity ($\rho$) transition around ~250 K ($T_p$, **Fig. 4a**), which is further confirmed through the specific heat measurements as demonstrated in **Fig. S3.** It exhibits typical semiconductor behavior above 250 K, while below 250 K, it assumes a metallic character with positive temperature dependence slope ranging from 175-230 K. The $\rho(T)$ curve is almost temperature independent between 25 and 200 K, the origin of which will be discussed in details in the theoretical calculation section. The ln$\rho$-$T^{-1/n}$ plot roughly follows a linear fitting for *n* = 4 at higher temperature region (**Fig. 4d** inset), suggesting a 3D Mott's variable-range-hopping dominated transport mechanism.[45]

As shown in **Fig. 4b**, the semiconducting HP-ReO$_3$ converts to a completely new metallic state, and the $T_p$ increases from 250 to 280 K at 0.7 GPa. With increasing pressure to 10.2 GPa, $T_p$ further decreases from 280 to 230 K. However, at higher pressure up to 20.2 GPa, $T_p$ becomes temperature independent. *In situ* HP Raman spectroscopy (**Fig. S4**) technique was used to study the vibrational properties of the materials under high pressure conditions. The absence of significant changes in the Raman spectra indicates that the structure can be maintained up to 38.9 GPa. The conductivity of HP-ReO$_3$ under different magnetic fields was measured at 0.7 GPa as shown in **Fig. 4b**. As the applied magnetic field increases from 0 to 9 T, the resistance of the material decreases, showing slightly negative MR. However, as the magnetic field increases from 6 to 9 T, the MR effect appears to be approaching saturation, which is attributed to its peculiar Fermi surface.[34]

*Magnetic properties*

The temperature-dependence $1/\chi(T)$ of HP-ReO$_3$ measured at field of $H$ = 5000 Oe is displayed in **Fig. 4c-d**. HP-ReO$_3$ manifests a typical paramagnetic behavior with no significant magnetic transition between 2 and 300 K. However, between 100 and 300 K, curvature of the $1/\chi(T)$ are not linear, and the blue auxiliary line deviates significantly from $1/\chi(T)$ after 250 K, which further evidences the presence of a structural phase transition near 250 K. The curves were fitted with a simple Curie formula: $\chi = \chi_0 + C/(T - \theta)$. The Curie-Weiss (CW) temperatures ($\theta$) and Curie constants ($C$) are: $\theta$ = -554 K and $C$ = 0.047, respectively. The fitting allowed us to extract the value of the effective magnetic moment $\mu_{eff}$ = 0.615 $\mu_B$, which is smaller than the calculated magnetic moment $\mu_{eff}$ = 1.41 $\mu_B$ since they are derived from an integration of the spin density over a confined sphere around the nuclei, whereas the fraction of electrons spilling out from these spheres is lost.

## *3.3 Theoretical Calculations*

Theoretical calculations show that the FM ground state is favored over the AFM one by 0.15 meV per unit cell with a weak magnetism (0.0015 $\mu_B$ per unit cell), which is consistent with experimental results. DFT-calculated spin-polarized band structure, presented in **Fig. 5a**, shows that the bands near Fermi level are mainly contributed by Re. The O atoms, on the other hand, contribute the most to the valence bands about 2 eV below the Fermi energy. The gap of 472 meV separates the bands contributed by Re and O atoms, indicating weak coupling between the *d*-orbital of Re atoms and the *p*-orbital of O atoms. Therefore, the electronic conductance in ReO$_3$ should be attributed to the electrons from the *d*-orbitals of Re atoms. Interestingly, our findings show that the conductance and valence bands touch at the *K*-point with a linear dispersion crossing the Fermi level, suggesting the possibility of a Dirac semimetal. The linear dispersion is basically contributed by the $d_{z^2}$ orbital of Re atoms, which differs from the well-known 3D Dirac semimetal Cd$_3$As$_2$ whose Dirac cone is contributed by the *p* orbital of As atoms.[46] According to the DFT results, the Fermi velocity is estimated to be $3.23 \times 10^6$ m/s, implying massless fermion and good electronic transport property in ReO$_3$. However, the Fermi velocity is about 1/3 of that of graphene,[47] which could be attributed to the heavier mass of Re atoms. Furthermore, we performed Fermi surface calculation, as shown in **Fig. 5b**. The Fermi surface displays $D_6$ rotation symmetry along $k_z$, which is consistent with the crystal symmetry. Focusing on the $k_z = 0$ plane. the *K* point is surrounded by the Fermi surface, corresponding to the linear dispersion at the *K* point, as is discussed above. Recently, Chen *et al*. claimed that the system is a Weyl semimetal if SOC is included, while an AFM ground state is required by the $C_2T$ symmetry. [48] Moreover, the situation in materials could be more complex than theoretical model. Therefore, further studies are desired to verify its topological properties.

In order to gain insight into the independence of resistance on temperature from 2 to 170 K, we calculated the electronic conductivity of $ReO_3$ in $P6_3$/mmc space group. Since the transition point of the resistance coincides with the phase transition point, we believe that this phenomenon may be related to the different structures. According to the experiment results, the O atoms of $ReO_3$ in the HP-phase possess fractional occupation, which implies the disorder of O atoms. Disorder will lead to an increase in scattering centers in the material and hinder the movement of electrons,[49] and even turn metals into insulators, as Anderson Localization reveals.[50] Therefore, the electronic conductivity should be negatively correlated with temperature. At the same time, semiconductors show a positive correlation between electron conductivity and temperature. The two cancel each other, so the electronic conductivity does not vary with temperature.

To verify this hypothesis, we perform theoretical calculations using a 3 × 3 × 3 supercell and consider ordered ($P6_322$), *ab* plane disordered, and disordered in all three dimensions ($P6_3$/mmc) conductance models, respectively. The results demonstrate that the conductivity increases (resistance decreases) with temperature for the ordered model, as shown in **Fig. 6**, consistent with the experimental findings presented in **Fig. 4a**. However, with increasing degree of disordering, the conductivity hardly changes with temperature. Notably, the experimental data reveal a straight-line relationship between conductivity and temperature within 2-170 K, while the platform observed between 170 and 230 K may be attributed to partial inhomogeneity of the material. These findings provide valuable insights into the electronic properties of HP-ReO$_3$ and the influence of structural disorder on electronic conductivity.

## 4. Conclusion

Investigation and synthesis of the high-pressure ReO$_3$ ($P6_322$) phase provide important insights into the distinct electronic and structural behavior of transition-metal honeycomb compounds. We have recorded the phase transition from $P6_322$ to $P6_3/mmc$, which occurs below 250 K due to the bifurcation of the oxygen position in the *ab* plane. Band structure calculation suggests that the high-pressure $P6_322$ display a signature of Dirac semimetal properties. This phenomenon is visibly reflected in its impact on paramagnetism and conductivity. The temperature-independent conductivity behavior observed below 250 K at ambient pressure, supported by theoretical calculations, suggests that disorder plays a significant role in the electronic response of this compound. The *in situ* pressure- and temperature-dependent resistance data show that O split can be inhibited by small external pressure and the $T_\text{p}$ varies irregularly within a small range as the pressure increases from 0.7-20.2 GPa. In summary, the multifaceted traits of the $P6_322$ polymorph of ReO$_3$ at high pressure offer a promising path for future research in comprehending the intricate interplay of structural and electronic properties in transition-metal honeycomb compounds.

**Credit author statement**

**Yifeng Han:** Data curation; Formal analysis; Funding acquisition; Investigation; Visualization; Writing - original draft. **Cui-Qun Chen:** Software; Writing - original draft; Visualization. **Hualei Sun**: Formal analysis. **Shuang Zhao**: Formal analysis. **Long Jiang**: Formal analysis; Methodology. **Yuxuan Liu**: Formal analysis. **Zhongxiong Sun**: Formal analysis. **Meng Wang**: Formal analysis. **Hongliang Dong**: Formal analysis. **Ziyou Zhang**: Formal analysis. **Zhiqiang Chen**: Formal analysis. **Bin Chen**: Formal analysis. **Dao-Xin Yao**: Writing – review & editing, Supervision, Funding acquisition; Investigation; Methodology; Project administration; Resources. **Man-Rong Li**: Writing – review & editing, Supervision, Funding acquisition; Investigation; Methodology; Project administration; Resources.

**Declaration of competing interest**

The authors declare that they have no known competing financial interests or personal relationships that could have appeared to influence the work reported in this paper.

**Data availability**

Data will be made available on request.

**Acknowledgments**

This work was financially supported by the National Natural Science Foundation of China (NSFC-22090041, 22105228, 92165204, 11974432), the NKRDPC-2022YFA1402802,

the Guangdong Basic and Applied Basic Research Foundation (Grant No. 2022B1515120014), the Program for Guangdong Introducing Innovative and Entrepreneurial Teams (2017ZT07C069), and Shenzhen International Quantum Academy (Grant No. SIQSE202102). The authors would like to thank the National Supercomputer Center in Guangzhou.19

**Figures and Captions**

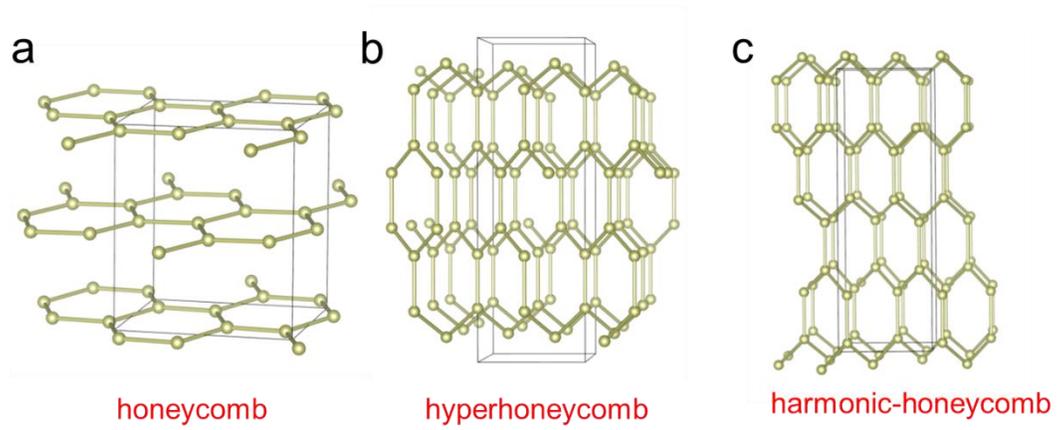

**Fig. 1.** Crystal structure of (a) honeycomb, (b) hyperhoneycomb and (c) harmonic-honeycomb lattice.

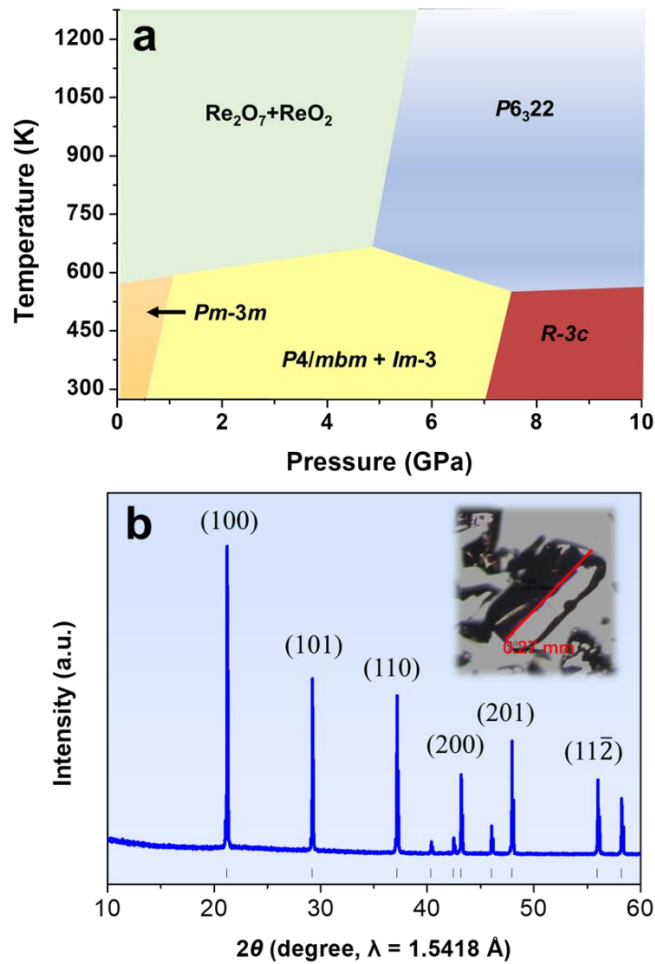

**Fig. 2. (a)** Pressure-temperature phase diagram of ReO$_3$; **(b)** Room temperature PXD pattern of HP-ReO$_3$ prepared at 7 GPa and 1173 K in hexagonal structure (*P*6$_3$22). Inset shows the image of one of the crystals grown at HPHT. Black ticks indicate the peak position of HP-ReO$_3$.

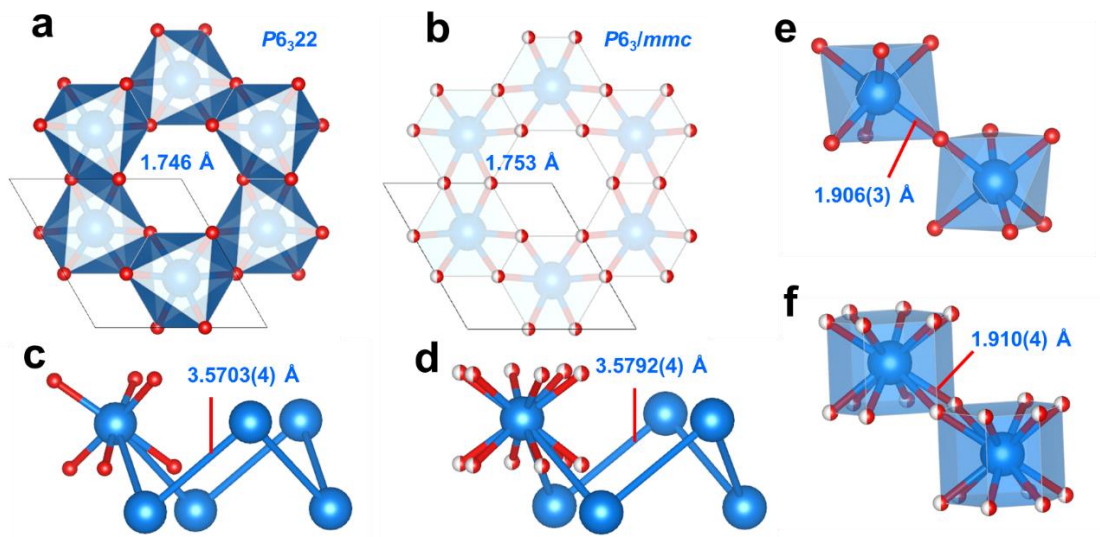

**Fig. 3.** Crystal structure of HP-ReO$_3$ at (**a, c, e**) 293 K and (**b, d, f**) 200 K. (**a-b**) crystal structure viewed alone the *c*-axis, black line indicates the unit cell dimension; Local structure of the (**c-d**) Mobius-like Re-Re circle and (**e-f**) ReO$_6$ polyhedron pair. Re, blue spheres; ReO$_6$ octahedra, blue; O, red spheres.

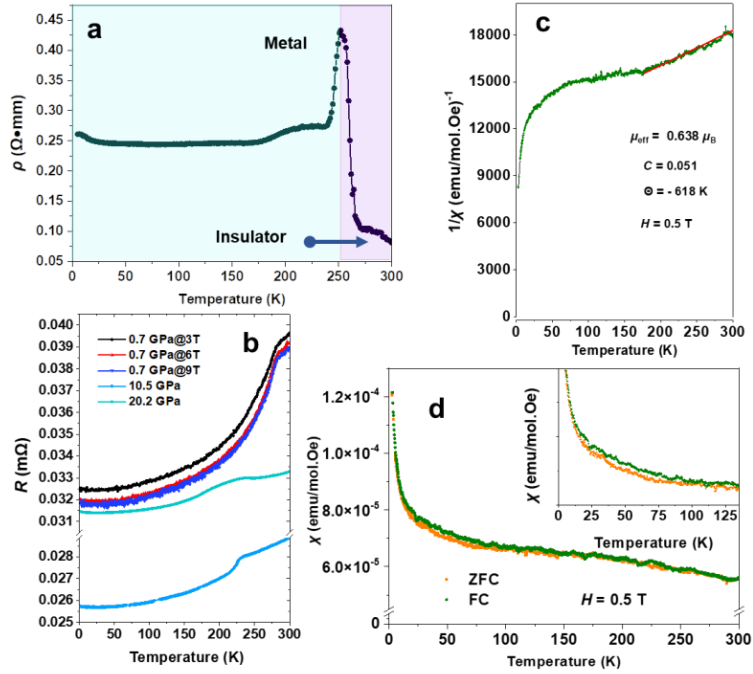

**Fig. 4**. Physical properties of HP-ReO$_3$. (**a**) Resistivity measurement of HP-ReO$_3$ at atmospheric pressure, insets show the plots of ln$\rho$ versus $T^{-1/4}$; (b) Temperature dependence of magnetoresistance at 0.7 GPa and resistance at 10.5 and 20.2 GPa; (c) The susceptibility inverse (1/$\chi$) versus temperature, the red line display the CW fitting; (d) Temperature-dependent magnetization ZFC and FC curves at 0.5 T.

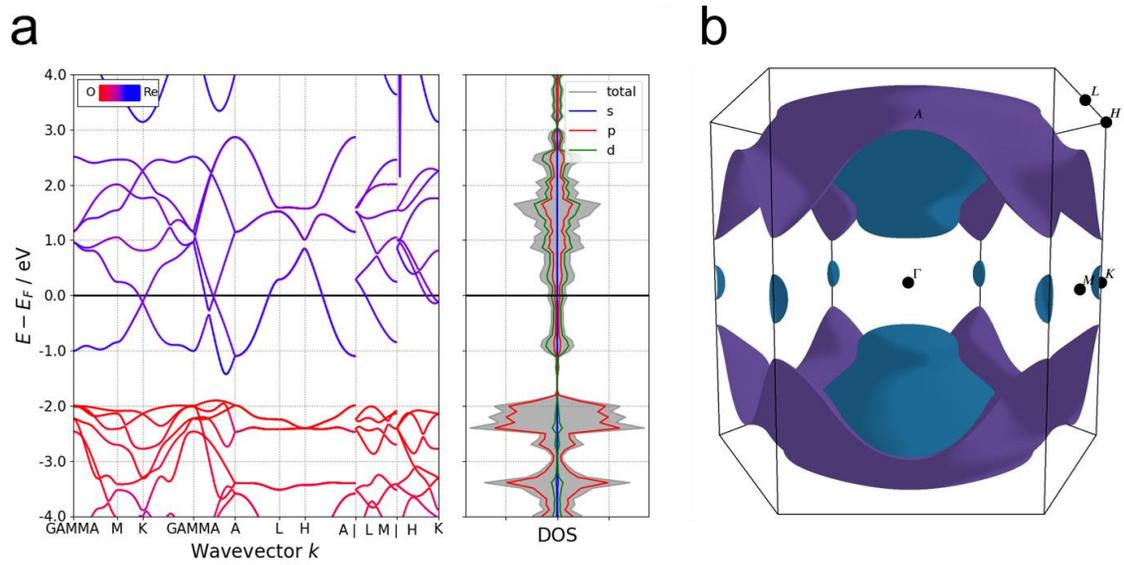

**Fig. 5.** (a) Calculated band structure of HP-ReO$_3$ in $P6_322$ space group. The blue bands are basically contributed by Re atoms while the red ones by O atoms; (b) Fermi surface of $P6_322$-ReO$_3$ in the first Brillouin zone.

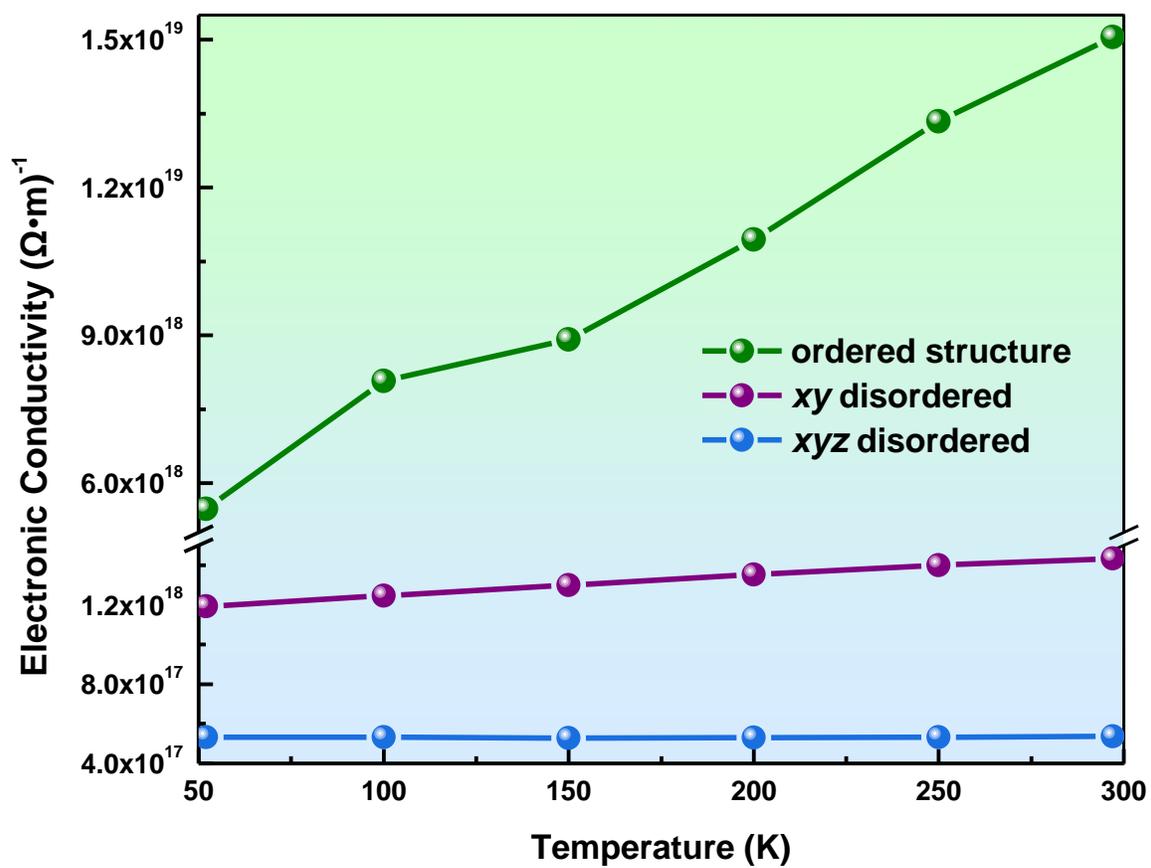

**Fig. 6.** Electronic conductivity of ReO$_3$ at different ordered states from theoretical calculations.